\documentstyle[prl,aps]{revtex}
\begin{document}

\newcommand{\be}{\begin{equation}}
\newcommand{\ee}{\end{equation}}
\newcommand{\bea}{\begin{eqnarray}}
\newcommand{\eea}{\end{eqnarray}}
\newcommand{\vp}{\varphi}

\draft

\title{Baryogenesis Below The Electroweak Scale}


\author{Lawrence M. Krauss[1,2] and Mark Trodden[1]}

\address{[1] Particle Astrophysics Theory Group,
Department of Physics, [2] Department of Astronomy \\
Case Western Reserve University, 
10900 Euclid Ave., Cleveland OH 44106-7079, USA.}

\date{\today}

\wideabs{

\maketitle

  
\begin{abstract}
\widetext

We propose a new alternative for baryogenesis which resolves a number of
the problems associated with GUT and electroweak scenarios, and which
may allow baryogenesis even in modest extensions of the standard model.  
If the universe never
reheats above the electroweak scale following inflation, GUT baryon
production does not occur and at the same time thermal sphalerons,
gravitinos and monopoles are not produced in abundance.  Nevertheless,
non-thermal production of sphaleron configurations via preheating could
generate the observed baryon asymmetry of the universe.
\end{abstract}


}

\narrowtext

The past twenty years have witnessed a roller-coaster ride as far as the
possible microphysical explanation of the observed
baryon asymmetry of the universe is concerned.  Before Grand Unified
Theories (GUTs), there were no physical theories which satisfied
Sakharov's three criteria for baryogenesis.  Subsequently, the simplest GUTs
were demonstrated to be able to account for the observed baryon to
photon ratio in the Universe today \cite{GUTs}.  Although proton
decay experiments soon ruled out the simplest theories, GUT
baryogenesis still remained a viable possibility in more complicated
models. However, GUTs also produced
several cosmological problems, the most urgent of which, the monopole problem,
led to the development of inflationary models for the early universe. 
However, while inflation does a good job of getting rid of monopoles, it
also gets rid of baryons.  Thus, unless the reheating scale following
inflation is large, standard GUT baryogenesis is impotent. This however,
raises the possibility of unacceptable defect production after inflation.
In addition, in SUSY models a high reheat temperature can result
in overproduction of gravitinos.
\cite{gravitinos}

Following these developments, it was recognized almost a decade later
that the standard electroweak model has the seeds for potentially viable
baryogenesis at the much lower electroweak scale ($\sim 10^2$ GeV).  
Coherent configurations
of gauge and higgs fields, first pointed out by 't Hooft \cite{tHooft76}, 
can violate baryon
number via non-perturbative physics.   At zero temperature this effect is
exponentially suppressed by the energy of a field configuration called the
{\it sphaleron}, and is essentially irrelevant.  However, as pointed
out by Kuzmin, Rubakov and Shaposhnikov \cite{krs}, and later discussed by
Arnold and McLerran \cite{a&m}, at finite temperature, sphaleron
production and decay can be rampant.  This has the virtue of allowing
unfettered baryon number violation.   Unfortunately, this can also be a
curse.  If the universe remains in thermal equilibrium until sphaleron
production ceases, the net effect of these processes will be to drive the
baryon number of the universe to zero, unless careful precautions are
made to ensure either out of equilibrium sphaleron decay, or quantum
number restrictions which forbid the elimination of the net baryon number.

Moreover, it has become clear that the Standard Model must be
supplemented by new fields at the weak scale to allow for
baryogenesis.  While certainly possible, this reduces one of the 
attractions of this idea.  

Thus, thermal sphaleron production creates both challenges and opportunities 
for the generation of the baryon asymmetry.  While it can wipe out any 
baryon number generated at the GUT scale, it offers the possibility
of electroweak baryogenesis, although in practice this is quite difficult to
achieve.

At the same time, the past few years have seen a revolution in thinking on
the subject of {\it reheating} after cosmological inflation. Careful
studies of the inflaton dynamics have revealed the possibility of a
period of parametric resonance, prior to the usual scenario of energy
transfer from the inflaton to other  fields. This phenomenon, which is
characterized by large amplitude, non-thermal excitations in both the
inflaton and coupled fields, has become known as {\it preheating}
\cite{KLS 94,STB 95}.

The new understanding of post-inflationary dynamics has seen applications
in a number of different phenomena. In particular, preheating has been
used to construct a new model of Grand Unified (GUT) baryogenesis \cite{KRT 98}
and to demonstrate how topological defects may be produced after inflation
even when the final reheat temperature is lower than the symmetry breaking
scale of the defects \cite{KLS 94,KKLT}. The question of this non-thermal
symmetry restoration has recently been argued to depend sensitively on
the expansion rate of the universe during reheating \cite{P&S 98}.

In this letter, we combine all of these ideas to present what we believe
is a viable, and attractive alternative which obviates many of the
problems with both standard GUT, and electroweak baryogenesis.  We
suggest that if inflation ends with reheating below the electroweak
scale, then a new route to baryogenesis may be possible.

  Our mechanism makes use of the ideas of non-thermal defect
production applied to the generation of gauge and Higgs field configurations
carrying non-zero Higgs winding number. Traditional models of electroweak
baryogenesis involve the motion of bubble walls during a strongly first 
order phase transition (for reviews see \cite{reviews}). 
The idea is that out of equilibrium sphaleron 
processes occur as the bubble walls sweep through space, and that CP
violation leads to a net baryon asymmetry being produced by these decays.
The question we address here is what happens if the reheat temperature 
after inflation is so low that there is no electroweak phase transition? 
Might the baryon asymmetry of the universe still be generated through 
electroweak physics?

Our fundamental observation is that, if topological defects can be
produced during preheating, then so can coherent configurations of gauge and 
Higgs fields, carrying nontrivial values of the Higgs winding number

\be
N_H(t) = \frac{1}{24\pi^2} \int d^3x\, \epsilon^{ijk} \hbox{Tr}
[U^{\dagger}\partial_iUU^{\dagger}\partial_jUU^{\dagger}\partial_kU] \ .
\label{higgswinding}
\ee
In this parameterization, the $SU(2)$ Higgs field $\Phi$ has been expressed as
$\Phi = ({\sigma}/{\sqrt{2}}) U  $, where $\sigma^2 =
2\left(\vp_1^*\vp_1 + \vp_2^*\vp_2 \right) = {\rm Tr} \Phi^\dagger
\Phi$,  and $U$ is an $SU(2)$-valued matrix that is uniquely defined
anywhere $\sigma$ is nonzero. 

These winding configurations are not stable and evolve to a vacuum 
configuration plus radiation. In the process fermions may be 
anomalously produced. If the fields relax to the vacuum by changing the Higgs 
winding then there is no anomalous fermion number production. However, if 
there is no net change in 
Higgs winding during the evolution (for example $\sigma$ never
vanishes) then there is anomalous fermion number production.
Since winding configurations will be produced
out of equilibrium (by the nature of preheating) and since CP-violation 
affects how they unwind, all the ingredients to produce a baryonic 
asymmetry are present (see \cite{LRT 97} for a detailed discussion of
the dynamics of winding configurations).

If the final reheat
temperature is lower than the electroweak scale, then then production of
small-scale winding configurations by resonant effects is analogous
to the production of local topological defects. In fact, the configurations 
that are of interest to us can be thought of as {\it gauged textures}.

To quantify this, we turn to recent numerical simulations
of defect formation during preheating \cite{KKLT,P&S 98}. While the number
density of defects produced has not been quantified, their existence has 
been verified. Thus, in order to get a rough underestimate of the number 
density, we may count defects in the simulations. Defect production has 
been studied in several different contexts. What is of interest to us here
is the case in which the symmetry breaking order parameter is not the
inflaton itself, but is the electroweak $SU(2)$
Higgs field, and is coupled to the inflaton \cite{Baacke}. Further, we are
interested in the number density of defects directly after preheating, since 
in the case of an $SU(2)$ order parameter, we expect the winding configurations
to decay very quickly, and so defect-antidefect pairs will not typically
have time to find each other and annihilate.  Finally, since the Higgs
winding is the only non-trivial winding present at the electroweak scale,
it is reasonable to assume that any estimates of defect production in 
general models can be quantitatively carried over to estimate of the
relevant Higgs windings for preheating at the electroweak scale.

Before attempting to give an estimate of the baryon asymmetry our mechanism 
produces, we would like to give an example in which the production of
winding configurations we require should occur at the necessary epoch. A simple
and natural implementation of our scenario can be found in {\it supernatural
inflation} models \cite{supernatural} of hybrid inflation \cite{hybrid}.
As a definite example, consider a two-field, flat direction hybrid inflation
model, with potential

\be
V(\phi,\psi)=M^4 \cos^2 \left(\frac{\phi}{\sqrt{2}f}\right) +\frac{1}{2}
m_{\psi}^2 \psi^2 + \frac{\lambda^2}{4}\psi^2 \phi^2 \ ,
\ee
where $M$, $f$, and $m_{\psi}$ are mass scales, and $\lambda$ is a 
dimensionless coupling. Note that in this case none of these fields
represents the electroweak Higgs field, nor will be be concerned about
preheating of the $\psi$ field, into the $\phi$ field, which may, or may
not occur, depending upon the parameter choices (for a detailed
discussion of preheating in hybrid models, see \cite{bellind}). Further,
for simplicity, we'll choose
$m_{\phi}\sim m_{\psi}
\sim m_{3/2} \sim 1$ TeV, although this is not crucial to the model.
In order to obtain an appreciable number of
e-foldings in this model, we must impose 

\be
M^4 > m_{\psi}^2 M_p^2 \ ,
\ee 
where $M_p \sim 10^{19}$GeV is the Planck mass. With our choice for 
$m_{\psi}$, this implies
$ M > 10^{11} \ {\rm GeV} $.
Note that near $\phi=0$, we may approximate $m_\phi \simeq M^2/f$, and 
therefore, for self-consistency we choose $f \sim M_p$.
Now, if, again for simplicity, we forbid direct Yukawa couplings of the
inflaton in the  superpotential, then what remains is a one loop term

\be
{\hat {\cal O}} \sim \frac{g^2}{\langle\phi\rangle}\int d^4\theta \, 
\chi^{\dagger}\chi\phi 
\ee
coupling the inflaton to electroweak superfields \cite{supernatural}. In this 
case, the final reheat temperature in this model is given by

\be
T_{RH} \sim {g}^{2/3}m_{\phi}^{5/6}M^{1/6} \ ,
\ee
For our purposes,  we will impose that the reheat temperature should
be insufficient to  allow thermal symmetry restoration in the electroweak
model. This ensures that any baryons produced will not be erased by
equilibrium  sphaleron processes. This condition reads

\be
g \sim 1.1\times 10^{-3} \ .
\ee
which is not an unnatural constraint.  Note that this constraint can
be weakened slightly if we allow $m_{\phi} $ to be less than 1 TeV.

Now, we are interested in whether parametric resonance into electroweak
fields occurs in this model. With the coupling of $\phi$ to the
electroweak fields given above, the condition for this to happen is
\cite{klebtkach}

\be
q = \frac{g^2 \phi_0^2}{2m_{\phi}^2} > 10^3 \ ,
\ee 
where $\phi_0$ is the value of $\phi$ at the end of inflation.
Since 
$\phi_0 \gg m_{\phi} \sim 1$ TeV, this condition is simple to arrange for 
the value of $g$ quoted above.  Note also that this model can be further
constrained in order to produce acceptable density fluctuations today. 
While we are merely presenting it as an example which accommodates our
mechanism, it is worth noting that the requirement to produce an
acceptable level of density fluctuations
\cite{supernatural} suggests
$\lambda \approx 10^{-4}$ for the range of the other variables
chosen here.  This constraint seems to be independent of the constraints
on parametric resonance and reheating of interest here, which depend upon
$g$ rather than $\lambda$.

It is also worth demonstrating here that even within 
the context of one field inflation models this mechanism can occur,
although a fine tuning seems to be required. In this case, the role of
the electroweak Higgs is explicit, however. This can be seen in an
extension of the model used in
\cite{P&S 98}. These authors studied domain wall production in a
chaotic inflation model with inflaton $\phi$, wall-field $\chi$ and
potential

\be
V(\phi,\chi) = \frac{1}{2} m^2 \phi^2 + \frac{1}{2}g^2\phi^2 \chi^2
+ \frac{1}{4} \lambda (\chi^2 - \chi_0^2)^2 \ ,
\ee
where $\chi_0$ is the symmetry breaking scale, $m$ is the inflaton mass, and
$\lambda$ and $g$ are dimensionless constants. Parametric resonance
occurs in this model \cite{P&S 98} if (i) $\lambda \chi_0^2 > g^2 \phi_0^2$, 
and (ii) $g^2 \chi_0^2 \ll m^2$, where $\phi_0 \simeq 0.2 m_{pl}$. In addition,
we know that the reheat temperature in this model is roughly bounded by

\be
T_{RH} \leq 10^{-3} (m\phi_0)^{1/2} \ .
\ee
Requiring again that any baryons produced not be erased by equilibrium 
sphaleron processes implies that (iii) $T_{RH} < \chi_0$.

Now, consider the above model with $\chi_0 = 250$ GeV, and $\lambda = 
{\cal O}(1)$, the values of the electroweak theory.  Choosing

\bea
g^2 & \ll & \min \left[ \frac{m^2}{\chi_0^2},\ 
\frac{\lambda \chi_0^2}{\phi_0^2} \right] \nonumber \\
m & \simeq & 10^{-9} {\rm GeV} \ ,
\eea
satisfies all the criteria above, and thus undergoes parametric resonance and
defect production.
Note that the parameter values required in this toy model are
not natural. However, the point of this example is merely to provide an
existence proof which makes explicit the constraints on such a
possibility.

While the authors of \cite{P&S 98} argue
that the generation of topological defects is suppressed during preheating when
the expansion of the universe is taken into account, we point out here that
at the electroweak scale it is a good approximation to consider the 
non-expanding case, in which defect production appears to be copious.

Based on the simulations of Khlebnikov {\it et al.} we see that, for
sufficiently  low symmetry breaking scales, the {\it initial} number
density of defects produced is very high.  Here, by initial, we mean not
the extremely high number that is found during the oscillations of the
inflaton (since these configurations quickly vanish) but rather the
number seen after copious  symmetry-restoring transitions cease. One may
perform an estimate from the first frame of Figure~6. of reference
\cite{KKLT}. The box size has physical size $L_{phys} \sim 50 \eta^{-1}$
where $\eta$ is the symmetry breaking scale, and we have, for simplicity,
assumed couplings of order unity. In this box there are of order $N=50$
defects at early times. This provides us with a very rough estimate of
the number density of winding configurations:

\be
n_{\rm configs} \sim \frac{N}{L_{phys}^3} \sim 4 \times 10^{-4} \eta^3\ .
\ee

In 
order to make a simple estimate of the baryon number that our mechanism can 
produce, the second thing we need to know is how
CP-violation may bias the decays of these configurations so that
a net baryon excess is created.

The effect of CP-violation on winding configurations can be very 
complicated, and in general depends strongly on the shapes of the 
configurations \cite{LRT 97} and the particular type of CP-violation. 
Examples are the case when CP-violation arises due to either
a CP-odd phase between Higgs fields in the two-Higgs doublet model, or through
higher dimension CP-odd operators in the electroweak theory.
However, in either case, the situation we consider here, when out of 
equilibrium configurations are produced in a background low-temperature 
electroweak plasma most closely
resembles local electroweak baryogenesis in the ``thin-wall'' regime. Winding
configurations, or topological defects, are produced when non-thermal 
oscillations take place in a region of space and restore the symmetry
there. We imagine that the symmetry is restored in a region and, since
the reheat temperature is lower than the electroweak scale, as the region 
reverts rapidly to the low temperature phase, the winding configuration is
left behind. 
In the absence of CP-violation in the coupling of the inflaton to the
standard model fields, we expect
a CP-symmetric ensemble 
of configurations with $N_H=+1$ and $N_H=-1$ to be produced.
(By this we mean that the probability for finding
a particular $N_H=+1$ configuration in the ensemble
is equal to that for finding its CP-conjugate $N_H=-1$ configuration.)
Then, without electroweak CP- violation, for every $N_H=+1$
configuration which relaxes in a baryon producing fashion
there is an $N_H=-1$ configuration which produces anti-baryons, and no
net baryogenesis occurs.
With CP-violation there will 
be some configurations which produce baryons whose CP-conjugate 
configurations relax to the $N_H=0$ vacuum without
violating baryon number.

While an analytic computation of the effect of CP-violation does not exist 
\cite{LRT 97}, there exist numerical simulations (e.g. \cite{Moore}), from 
which one expects that the asymmetry in the number density of
decaying winding configurations should be proportional to a dimensionless
number, $\epsilon$, parameterizing the strength of the source of CP-violation.
Now, at the electroweak scale the entropy density is 
$s\simeq 2\pi^2 g_* T^3/45$, where $g_*\sim 100$ is the effective number
of massless degrees of freedom at that scale. Thus, the final baryon to
entropy ratio generated by our mechanism is

\be
\eta \equiv \frac{n_B}{s} \sim \epsilon \, g_*^{-1} 
\frac{n_{\rm configs}}{T_{RH}^3} \ .
\label{etafinal}
\ee
Plugging in the approximate numbers that we obtained earlier, this yields

\be
\eta \equiv \sim 10^{-6} \epsilon \ .
\ee
This is our final estimate.

This estimate while rough, suggests that the mechanism we are proposing
here could viably result in a phenomenologically allowed value
of $\eta \sim 10^{-10}$, with CP violating physics which is certainly
within the range predicted in SUSY models for example.

The advantages of such a mechanism are several, and we briefly summarize
them here: (1) No thermal sphaleron production subsequently takes place
to wash out any baryon number that is produced, (2) no excess production
of gravitinos or monopoles is implied, (3) 
a prohibitively large rates of proton decay is not implied, and (4) the
existence of a great deal of new physics near the electroweak scale is
not required. These reasons, combined with the phenomenologically
interesting estimate above, suggest such a mechanism should be explored
more closely in the future. A more complete analysis would involve a
numerical solution to the coupled
$SU(2)$-inflaton equations of motion, in the presence of CP-violation. We
expect to undertake such an analysis in a later publication.

We thank Matthew Parry, Richard Easther and Lisa Randall 
for useful discussions. This work was supported by the 
Department of Energy (D.O.E.).

\end{document}